\documentclass[showpacs,prb,aps,twocolumn]{revtex4}
\usepackage{graphicx}
\usepackage[normalem]{ulem}

\usepackage{amssymb}

\begin{document}

\title{Two-magnon Raman scattering in A$_{0.8}$Fe$_{1.6}$Se$_2$ systems:
competition between superconductivity and antiferromagnetic order}

\author{A. M. Zhang}
\author{J. H. Xiao}
\author{Y. S. Li}
\author{J. B. He}
\author{D. M. Wang}
\author{G. F. Chen}
\author{B. Normand}
\author{Q. M. Zhang}
\affiliation{Department of Physics, Renmin University of China, Beijing
100872, P. R. China}

\author{T. Xiang}
\affiliation{Institute of Physics, Chinese Academy of Sciences, Beijing
100190, P. R. China}
\affiliation{Institute of Theoretical Physics, Chinese Academy of Sciences,
P.O. Box 2735, Beijing 100190, P. R. China}

\date{\today}

\begin{abstract}

We have performed Raman-scattering measurements on high-quality single
crystals of A$_{0.8}$Fe$_{1.6}$Se$_2$ superconductors of several
compositions. We find a broad, asymmetric peak around 1600 cm$^{-1}$
(200 meV), which we identify as a two-magnon process involving optical
magnons. The intensity of the two-magnon peak falls sharply on entering
the superconducting phase. This effect, which is entirely absent in the
non-superconducting system KFe$_{1.5}$Se$_2$, requires a strong mutual
exclusion between antiferromagnetism and superconductivity arising from
proximity effects within regions of microscale phase separation.

\end{abstract}


\maketitle

\section{Introduction}

The investigation of iron-based superconductors experienced an exciting
breakthrough in the synthesis of arsenic-free, potassium-intercalated
FeSe.\cite{XLChen} This material has a maximum $T_c = 32$ K, which is
duplicated on replacing K by other monovalent ions, including Rb, Cs,
and Tl (denoted henceforth by A).\cite{AFeSe} These systems are found
to be non-stoichiometric both in A and in Fe content, and numerous studies
have been performed which identify ordered structures of the vacancies in
the Fe plane. Long-range magnetic order appears to be involved implicitly
in the selection of the ordered vacancy pattern, which leads directly
to the most anomalous property of the A$_{x}$Fe$_{2-y}$Se$_2$ systems, namely
the apparent coexistence of antiferromagnetism and superconductivity.

The initial suggestion of Fe vacancy order came from transmission electron
microscopy (TEM)\cite{TEM} and X-ray diffraction (XRD).\cite{Xray} Both
infrared\cite{Infrared} and Raman\cite{ZAM} measurements observed many more
phonon modes than would be expected for stoichiometric Fe superconductors,
such as the BaFe$_2$As$_2$ (``122'')\cite{Iliev} or Fe(Se,Te) ("11")
structures.\cite{QMZhang} These were shown\cite{ZAM,ZAM1} to be
consistent with the $\sqrt{5} \! \times \! \sqrt{5}$ Fe-vacancy pattern
expected for an Fe content of 1.6. This low-temperature structure was
then demonstrated conclusively by neutron diffraction studies of several
crystals.\cite{wbnd} However, other Fe-vacancy structures have also been
observed in A$_{x}$Fe$_{2-y}$Se$_2$, with neutron diffraction revealing not
only the $\sqrt{5} \! \times \! \sqrt{5}$ phase ($I4/m$) but also a
vacancy-disordered ($I4/mmm$) phase above 500 K and a $\sqrt{2} \! \times
\! \sqrt{2}$ phase ($Pmna$) below 500 K.\cite{wbnd2} Experiments by scanning
tunneling microscopy (STM), TEM, X-ray, and neutron scattering have suggested
the possibility that superconductivity may also be present in the $\sqrt{2}
\! \times \! \sqrt{2}$ phase.\cite{sqrt2}

Still more anomalous is that the ordered $\sqrt{5} \! \times \! \sqrt{5}$
Fe-vacancy structure is not only accompanied by, but to a significant extent
stabilized by, a very strong antiferromagnetic (AF) order.\cite{wbnd} Setting
in at a remarkably high transition temperature around 520 K, the spin order
consists of AF-coupled four-site ferromagnetic blocks, with large on-site
moments of 3.3$\mu_B$. The transition was confirmed by high-temperature bulk
magnetic measurements\cite{XHChen} and the large moment by M\"ossbauer
spectroscopy,\cite{Mossbauer} while muon spin rotation ($\mu$SR) experiments
put the magnetic volume fraction at 90$-$95\%.\cite{MuSR} In addition,
inelastic neutron scattering (INS)\cite{PCDai} and two-magnon Raman
measurements (below) also show the presence of a strongly magnetic phase.
However, nuclear magnetic resonance (NMR) experiments\cite{NMR} find singlet
superconductivity with only weak AF fluctuations, and no detectable magnetic
order. No evidence of magnetic order is present in the Fermi surfaces observed
by angle-resolved photoemission spectroscopy (ARPES).\cite{ARPES} Both of
these results suggest that the presence of magnetic order and superconductivity
in the A$_{x}$Fe$_{2-y}$Se$_2$ superconductors should be the result of a
macroscopic phase separation rather than any type of microscopic
coexistence or cohabitation.

The question of coexistence has recently been the focus of the most intense
experimental investigation. A clear consensus has emerged in favor of phase
separation, which has been reported in optical\cite{Charnukha} and ARPES
measurements.\cite{Chen} Specially designed M\"ossbauer,\cite{Ksenofontov}
X-ray,\cite{Ricci} and in-plane optical spectroscopic measurements\cite{Yuan}
have all been used to specify that the phase separation occurs on nanoscopic
length scales, and this has now been observed directly by STM on epitaxially
grown films.\cite{Li} Some of these authors\cite{Chen,Li} have further
identified that vacancy order is a property only of the AF phase, while
the superconducting phase is uniform and may be composed of stoichiometric
AFe$_2$Se$_2$ regions, a conclusion also suggested by INS.\cite{Friemel}
However, none of these works has shed any light on the coupling of the
two phases, by which is meant the key question of whether superconductivity
and antiferromagnetism can merely exist side by side, or whether they also
influence each other.

In this paper, we answer this question by using Raman scattering to
measure the high-energy excitations in three superconducting (SC)
A$_{0.8}$Fe$_{1.6}$Se$_2$ samples and one non-SC KFe$_{1.5}$Se$_2$ crystal.
We observe broad two-magnon scattering signals whose energy demonstrates
a common origin in the optical magnon modes of the complex unit cell.
Strikingly, the continuous increase of the two-magnon intensity with
decreasing temperature ends abruptly at $T_c$ in the SC systems, displaying
 a rapid drop at $T < T_c$. Our results demonstrate the intrinsic opposition
of superconductivity to bulk magnetism. They also add to the weight of
evidence that the size of the cohabitating regions of AF order and
superconductivity in the A$_{0.8}$Fe$_{1.6}$Se$_2$ materials is microscopic.

The structure of the manuscript is as follows. In Sec.~II we introduce the
materials and the experimental techniques employed. Section III presents
our primary results for two-magnon Raman scattering, discussing the broad
signal we observe in all samples, its origin, polarization-dependence, and
a fine structure of subpeaks. The temperature-dependence of our results is
so remarkable that it merits a separate section, Sec.~IV, where we quantify
the sharp drop below $T_c$ by means of intensity integrations. In Sec.~V
we discuss the consequences of our observations for models of cohabitation
between magnetism and superconductivity. A short summary can be found in
Sec.~VI.

\section{Materials and Methods}

The FeSe-based single crystals used in our measurements were grown
by the Bridgman method,\cite{GFChen} and were cleaved from exactly the same
batch as those used in neutron scattering experiments by Bao {\it et
al.}\cite{wbnd,wbnd2} The stoichiometry of each crystal is determined by
inductively coupled plasma atomic emission spectroscopy (ICP-AES) and by
neutron diffraction refinement.\cite{wbnd} Thus we identify our SC samples
as K$_{0.8}$Fe$_{1.6}$Se$_2$ ($T_c = 32$ K), Tl$_{0.5}$K$_{0.3}$Fe$_{1.6}$Se$_2$
($T_c = 29$ K), and Tl$_{0.5}$Rb$_{0.3}$Fe$_{1.6}$Se$_2$ ($T_c = 31$ K), while
the non-SC crystal is KFe$_{1.5}$Se$_2$. XRD analysis shows no discernible
secondary phase in any of the crystals. Their resistivities and magnetizations
were measured respectively with a Quantum Design physical properties
measurement system (PPMS) and the PPMS vibrating sample magnetometer (VSM).
Both quantities exhibit sharp SC and diamagnetic transitions for all three
samples (shown below, in the insets of Fig.~2), proving the high quality of
the crystals used in our Raman investigation. Magnetization measurements were
performed both before and after the Raman investigation, confirming that no
changes in the SC state occur during our experiments.

Before performing a high-energy Raman measurement, we first cleaved a
piece of crystal (approx.~$1\times2\times0.2{\rm mm}^{3}$) in a glove box
to obtain a flat, shiny $(ab)$-plane surface. The freshly-cleaved crystal
was sealed under an argon atmosphere and transferred rapidly into the
cryostat, which was evacuated immediately to $10^{-8}$ mbar. We use a
pseudo-backscattering Raman configuration with a triple-grating monochromator
(Jobin Yvon T64000). All of the Raman spectra were collected with a 532 nm
solid-state laser (Torus 532, Laser Quantum), except for the spectrum in
Fig.~1(b), where a 633 nm Melles Griot He-Ne laser was used to exclude
the possibility of photoluminescence effects (Sec.~III). The beam is
focused to a spot diameter of 20 $\mu$m at the sample surface and its power
reduced below 0.5 mW at low temperatures. The temperature in the spot is
calibrated from the Stokes and anti-Stokes spectra.

The T64000 spectrometer works in subtractive mode. The third-stage grating,
which has 1800 grooves per mm, has a resolution of approximately 0.6cm$^{-1}$.
Each spectral window covers 400-600 cm$^{-1}$. Because our spectrometer is not
equipped with a low-density grating, we measure a ``long'' spectrum of the
types shown in Secs.~III and IV  by combining several short spectral windows.
These windows can be joined smoothly because the stray-light level is
negligible at the high wavenumbers we probe. Long spectra from 600 to 3920
cm$^{-1}$ were obtained by combining seven spectral windows. The combination
quality is high due to the high quality of the cleaved crystal surface, which
allows a precise determination of the relative spectral intensity.

This procedure is accurate for the shapes of broad features. At times when
we found the intensities to have an error of a few percent, we took additional
steps to suppress their fluctuations. The most fundamental was to
repeat our measurements, which was particularly important around $T_c$
(Sec.~IV), and where we performed three or more sets of scans in order to
obtain a completely reproducible spectrum at each temperature. Another
step was to merge the high-energy tails, as these should depend only on
``extrinsic'' factors (such as the geometry, the surface, and the instrument)
rather than on intrinsic materials properties. If the tails did not merge
automatically, the whole spectrum was multiplied by an overall factor to
ensure a match. We also normalized the low-energy spectra using a
Bose-Einstein thermal factor, and this led to very close agreement between
spectral segments in almost all cases. Because the low-energy spectra lie
rather close to the laser line, they have a higher possibility of being
contaminated by stray light, and thus we took both the high- and low-energy
spectra as criteria to guarantee that each spectrum is collected with the
correct relative intensity.

\begin{figure}
\includegraphics[angle=0,width=8cm]{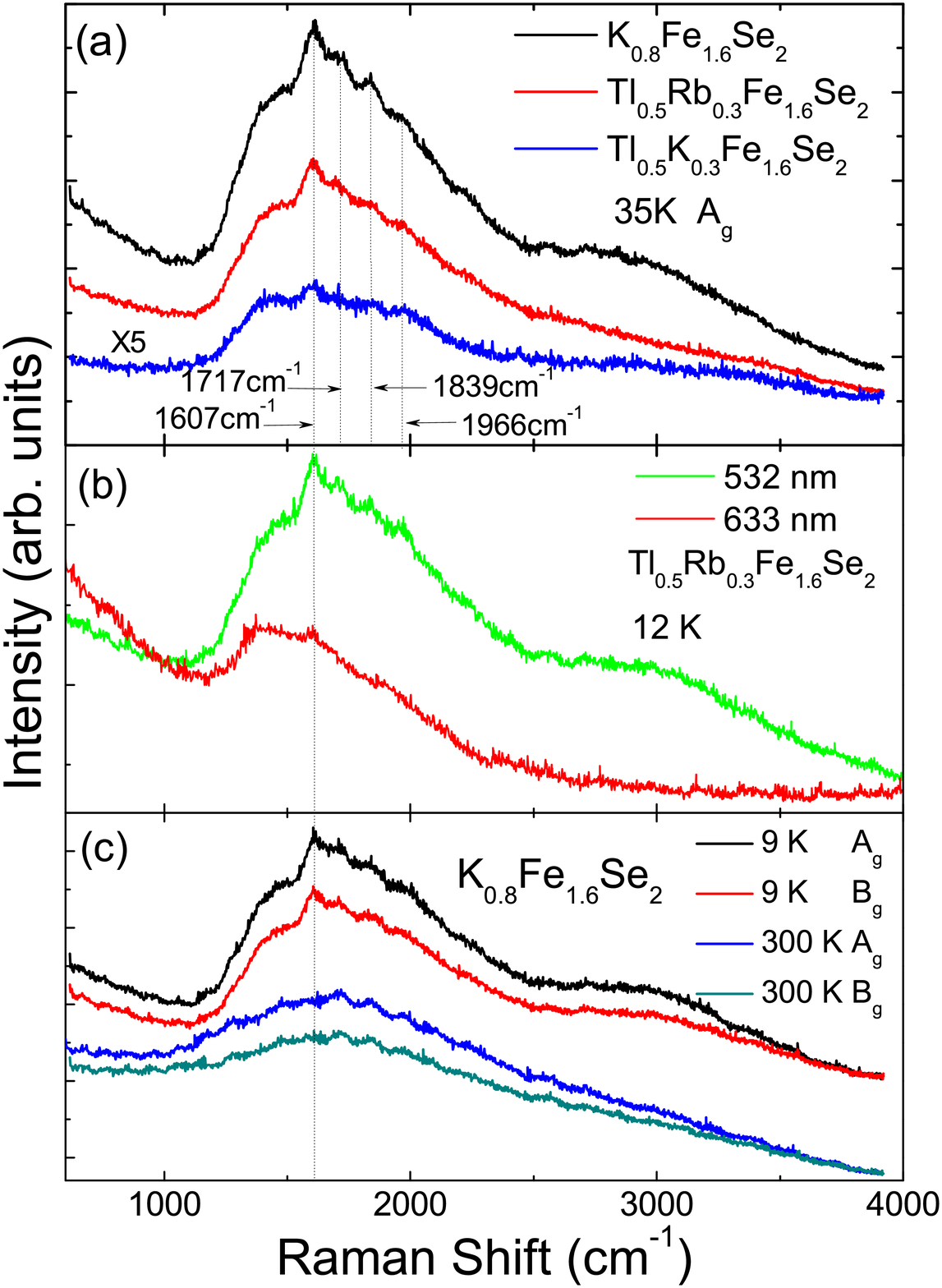}
\caption{(Color online) (a) Raman spectra for the three SC samples measured
at 35 K in the $A_g$ channel. Dotted lines mark peak frequencies observed
in all samples. (b) Comparison of Raman spectra measured with excitation
wavelengths of 633 nm (red) and 532 nm (green). (c) Similar two-magnon
spectra are observed in both $A_g$ and $B_g$ channels.}
\label{fig1}
\end{figure}

\section{Two-magnon Raman Scattering}

In Raman scattering by magnetic excitations, the dominant contribution
is given by two-magnon processes involving spin excitations of equal and
opposite momenta close to the zone boundary. This is the origin of the
well-defined peak at energies $\omega \approx 3J$ in square-lattice AF
structures,\cite{K2NiF4} which has been widely exploited in cuprate
superconductors despite a strong broadening of the signal due to AF spin
fluctuations.\cite{cuprate} Because the penetration depth for a Raman
measurement is up to 100 nm in typical systems with lower carrier densities,
this can be regarded as a bulk rather than a surface technique. Thus Raman
scattering plays an essential role in studying magnetism in correlated
electron systems. For Fe-based superconductors, the only two-magnon
measurements reported to date are for the 122 and 11 systems.\cite{Sugai}

\subsection{Sample characteristics}

Figure 1(a) shows the Raman spectra of the three SC samples, measured to
4000 cm$^{-1}$ at 35 K. All three samples show a broad and asymmetric peak
around 1600 cm$^{-1}$ (200 meV), while some also suggest a high-energy
shoulder at 2800$-$3500 cm$^{-1}$. The most remarkable feature is that all
three spectra are qualitatively identical, proving that the measured behavior
is truly intrinsic to the FeSe layers. While the signal intensities vary with
the A-site dopant, whose non-stoichiometry leads to clear disorder effects,
these are arbitrary units. As in any Raman measurement, they depend strongly
both on the sample (at fixed composition) and on the chosen spot on each
sample. For this reason, in the remainder of manuscript we focus on the
energetics of the scattering response, which are read from the Raman shift,
and on relative intensities only within a single measurement.

Before proceding further, in Fig.~1(b) we compare the spectra obtained in
one sample with two different excitation wavelengths, 633 nm and 532 nm.
While the background terms are clearly different, the broad feature around
1600 cm$^{-1}$ is certainly reproduced at 633 nm, albeit with a global
intensity reduction. In fact a similar reduction effect is also observed
in studying two-magnon processes in cuprates.\cite{Sugai1} This comparison
confirms that the 1600 cm$^{-1}$ feature is indeed a real Raman signal, and
not a photoluminescence effect. In contrast, the possible high-energy shoulder
around 3000 cm$^{-1}$ is not present when excited by a 633 nm laser, which
suggests very strongly that it is due to photoluminescence, rather than to
an intrinsic process.

\begin{figure}
\bigskip
\includegraphics[angle=0,width=7.0cm]{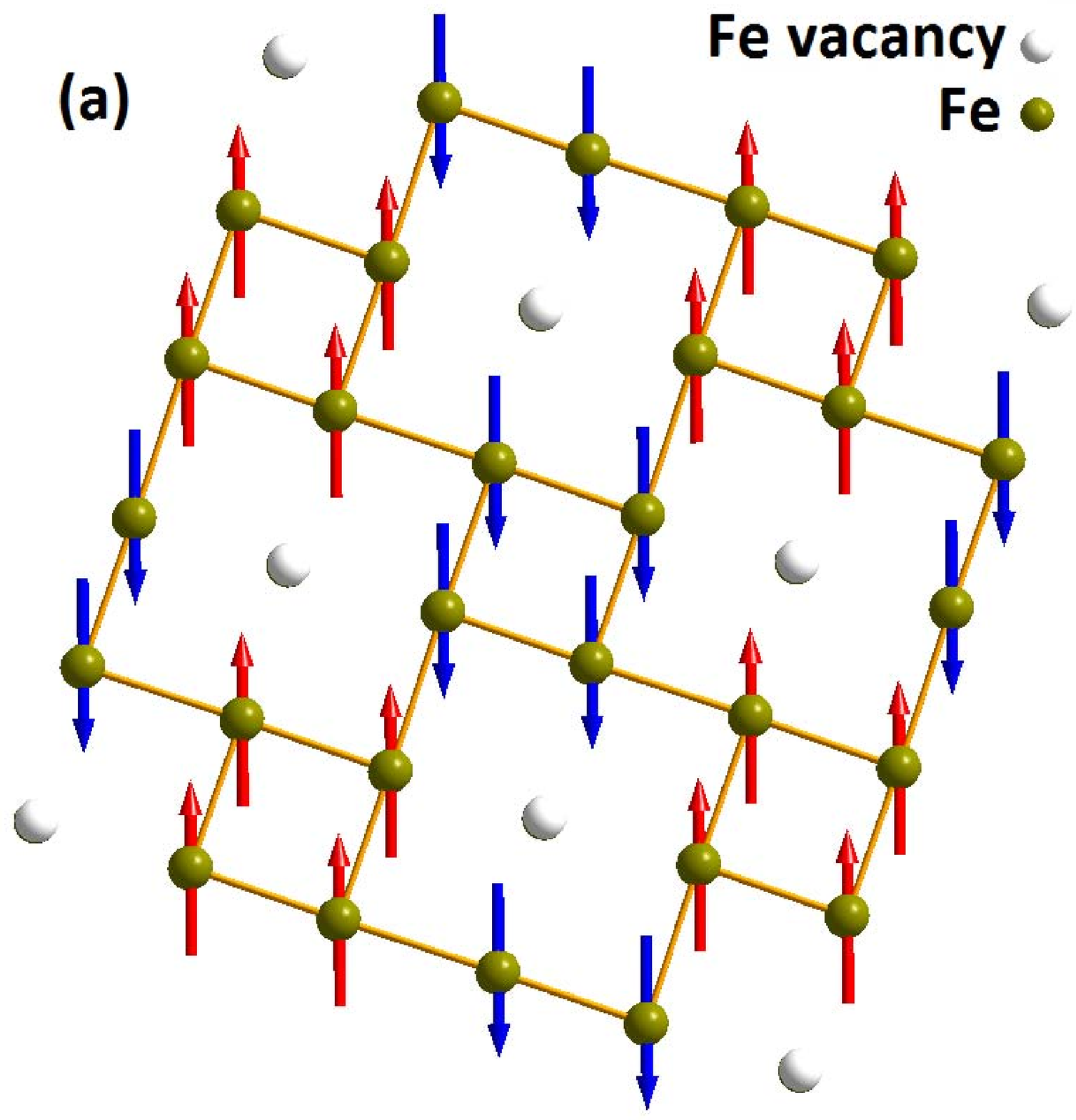}
\centerline{\phantom{(a)}}
\includegraphics[angle=0,width=8.0cm]{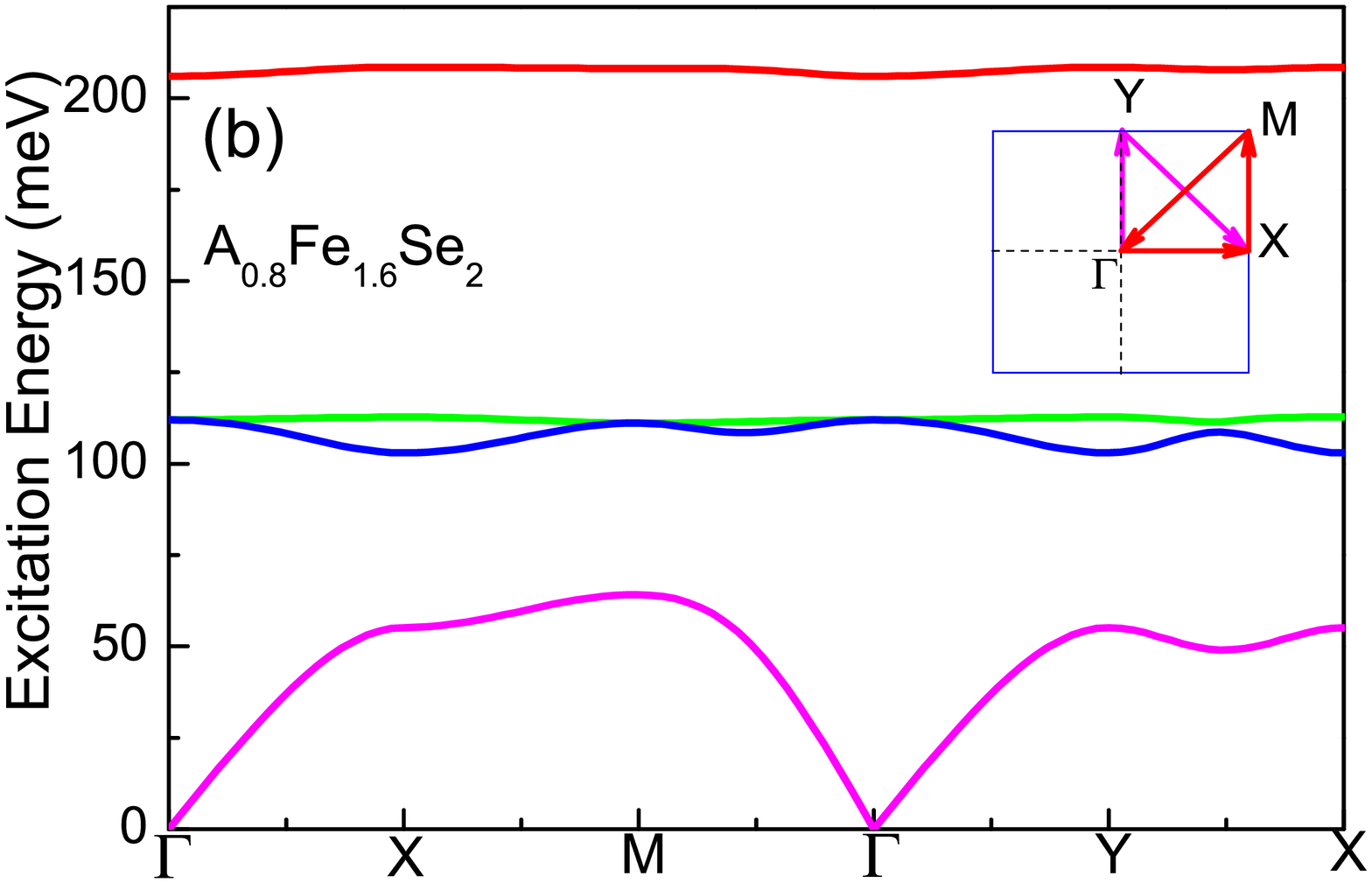}
\caption{(Color online) (a) Schematic representation of the $\sqrt{5} \!
\times \!\! \sqrt{5}$ vacancy-ordered structure and magnetically ordered
configuration of the Fe$_{1.6}$Se$_{2}$ plane. (b) Corresponding magnon band
structure; the four-site unit cell results in four magnon bands, of which
one is acoustic and the remaining three are optical. With this parameter
choice, which is based on the INS results of Ref.~[\onlinecite{PCDai}],
the dominant contributions to the two-magnon Raman intensity are due to
this optical modes at 110$-$125 meV (shown in blue and green); we caution
that some alterations to the parameter set may be required to account for
the absence in our measurements of a high-lying optical mode.}
\label{fig2new}
\end{figure}

We also comment briefly on a polarization analysis of our spectra. The
symmetry-dependence of the two-magnon signal is shown in Fig.~1(c). The
peaks measured in the $A_g$ and $B_g$ channels show similar lineshapes and
only small differences in intensity. This behavior is quite different from
the cuprate superconductors, where two-magnon processes are allowed only
in the $B_{1g}$ channel, and arises because the $\sqrt{5} \! \times \!
\sqrt{5}$ vacancy-ordered structure lowers the lattice symmetry of
the A$_{0.8}$Fe$_{1.6}$Se$_2$ materials. The presence of a two-magnon
signal in both channels has been demonstrated in local-moment models
for Fe-based superconductors in Ref.~[\onlinecite{rcjksd}].

\subsection{Optical magnons}

The origin of the primary spectum is indisputably magnetic: optical phonon
modes appear only below 300 cm$^{-1}$,\cite{Infrared,ZAM} and high-order
multi-phonon processes may be safely excluded. An electronic origin in
inter-band transitions is excluded by an absence of 200 meV features in
angle-resolved photoemission spectroscopy (ARPES).\cite{ARPES} To understand
the magnetic contributions, we consider\cite{ryyl} the magnon bands of the
$\sqrt{5} \! \times \! \sqrt{5}$ magnetic structure [Fig.~2(a)]. These bands
consist of one acoustic branch, which corresponds to processes involving the
effective block spin of each four-site unit, accompanied by three optical
branches arising from local processes on the square lattice of Fe sites. They
are shown schematically in Fig.~2(b), where the energy scales are motivated
by an inelastic neutron scattering (INS) study of the magnetic excitations
in Rb$_{0.89}$Fe$_{1.58}$Se$_2$.\cite{PCDai} This investigation suggested that
the acoustic branch extends to 70 meV, while two rather flat optical branches
are present at 110$-$120 meV, accompanied by a high-lying optical branch
around 200 meV.

In a local picture of two-magnon Raman scattering, the dominant energy
scale is given by reversing the spins on the shortest bond(s), in the
system. The magnetic exchange constants within the Fe plane estimated
in Ref.~[\onlinecite{PCDai}] give an energy of order 220 meV, on both nearest-
and next-neighbor bonds. This is exactly the peak energy we observe.
Such local processes are contained in the optical magnon bands, and
therefore our primary signal may be ascribed to two-magnon processes
involving the 110$-$120 meV optical magnons of Ref.~[\onlinecite{PCDai}].
One-magnon signals, which would be expected at 800$-$900 cm$^{-1}$, appear
to be small. We do not observe significant contributions from the acoustic
magnon branch. Our results also do not provide a reliable indication for
the presence of a high-lying optical magnon mode (the 3000 cm$^{-1}$ feature
reported by INS), and therefore we use the exchange constants of
Ref.~[\onlinecite{PCDai}] for illustration only. Finally, the breadth in
frequency of the signal we observe is strongly reminiscent of cuprates,
and it cannot be explained solely on the basis of multiple magnon bands;
it constitutes clear evidence for strong spin-fluctuation effects occurring
in the FeSe planes of the A$_{0.8}$Fe$_{1.6}$Se$_2$ materials.

\subsection{Fine Structure}

The two-magnon intensity [Fig.~1(a)] also contains several individual peaks
superposed on the broad, two-magnon signal, occurring at 1607, 1717, 1839,
and 1966 cm$^{-1}$. These apparent resonances are small in intensity and
rather narrow in energy. Their location and regular spacing are completely
reproducible between the different superconducting samples, adding strongly
to the evidence that all of the features of our observed response are
characterisic of the FeSe planes.

We do not yet have a complete explanation for these features. Although the
majority of our measured intensity is a two-magnon signal, in our calculated
Raman response we have not been able to reproduce these features on the
basis of the magnon band structure of Ref.~[\onlinecite{PCDai}]. Even mode
dispersions as flat as the magnon bands appearing between 110 and 125 meV
in Fig.~2(b) do not have density-of-states effects (arising from their upper
and lower band edges) sufficiently strong that peaks of this sharpness can
emerge. The regularity of the peak spacings is also difficult to reproduce.

Another possibility would be the ``two-magnon plus phonon'' processes
proposed to explain sharp features in the two-magnon Raman response of
cuprates.\cite{rls,rfs} By this mechanism, the phonon or phonons act(s) to
lower the symmetry of the net two-magnon absorption process, and its wave
vector allows contributions from many more pairs of magnon wave vectors. In
the context of cuprates, particularly strong additional contributions were
obtained from in-plane ($\pi$,0) phonons involving the O atoms. While analogous
processes in the FeSe plane are more difficult to isolate because of its
buckled structure (the Se atoms are located alternately above and below
the Fe plane), we do have, from our own previous results for Raman scattering
by phonons in K$_{0.8}$Fe$_{1.6}$Se$_2$,\cite{ZAM} a rather complete accounting
of the available vibrational modes. We found experimentally that the most
intense phonon modes lie at energies of 135, 203, and 265 cm$^{-1}$. All have
$A_g$ symmetry and we were able theoretically to show that all three involve
predominantly $c$-axis motions of the Se atoms. Further microscopic analysis
is required of whether different combinations of phonon excitations on the
intervening Se atom during the local spin-flip process can enhance the
creation of magnon pairs.

\begin{figure}
\includegraphics[angle=0,width=8.5cm]{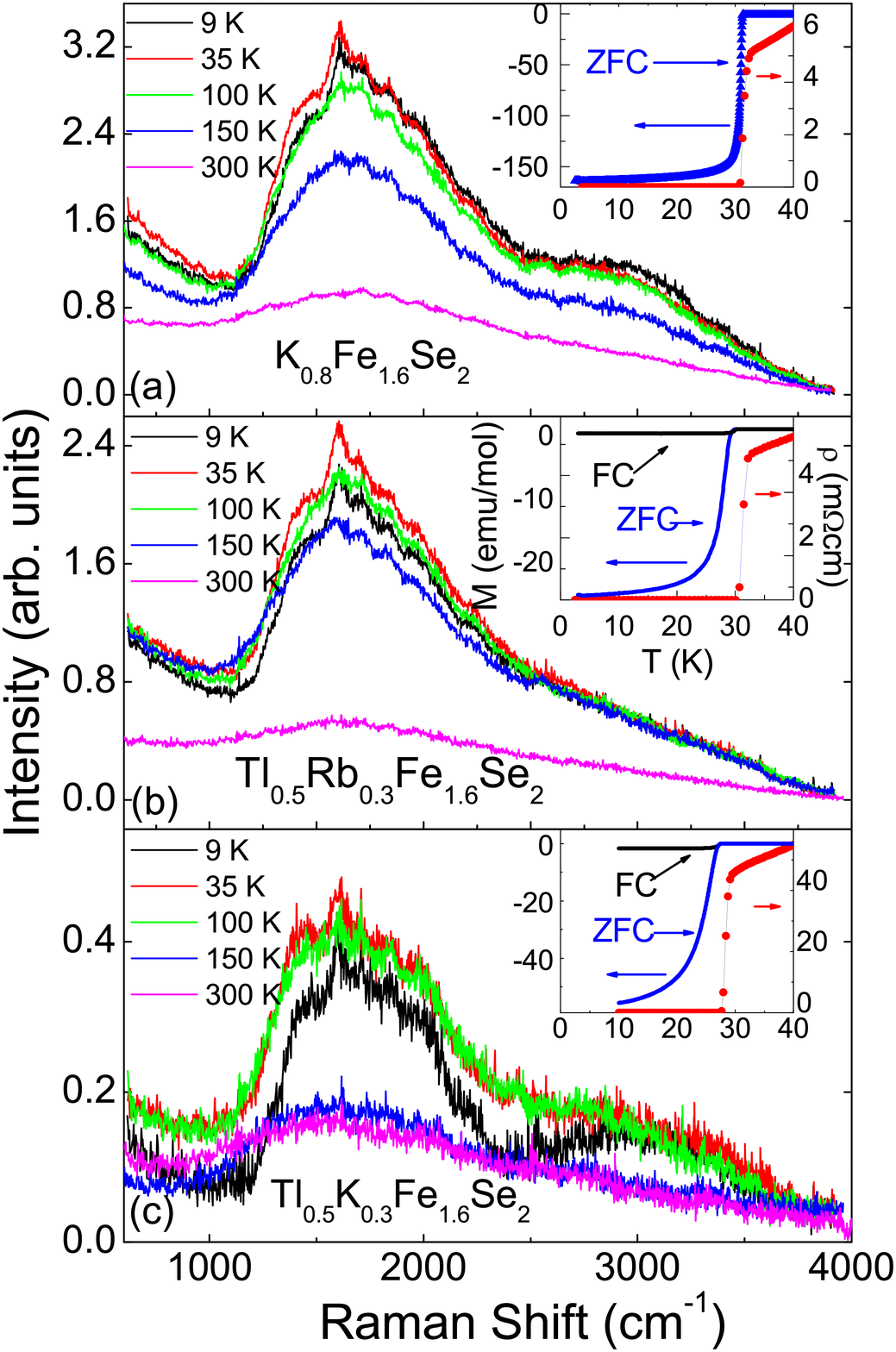}
\caption{(Color online) Two-magnon scattering intensity in the three SC
crystals at selected temperatures. Intensities at 35 K (red) are higher
than those in the SC state (black). Insets: resistivity and magnetization
data at and below $T_c$.}
\label{fig3}
\end{figure}

\section{Temperature-dependence}

In Fig.~3 we show the two-magnon signals for all three SC samples over
a range of temperatures. Lowering the temperature causes the peaks to become
sharper and their intensity to rise continuously, as expected from a reduced
thermal scattering of magnons. Remarkably, this tendency is interrupted at
$T_c$. For the spectra in Fig.~3(a), the intensity in the SC state (9 K)
clearly lies below that in the normal state near $T_c$ (35 K) at frequencies
below the peak, and then lies on or slightly above it. This behavior is not
ambiguous in Fig.~3(b), where the low-temperature signal lies below the
normal-state one everywhere. Finally, the difference is quite dramatic in
Fig.~3(c), although the data here are noisier.

To place this result in perspective, we show in Fig.~4 the corresponding
spectra for the non-SC sample, measured to 5300 cm$^{-1}$. While higher
temperatures suppress the intensity more strongly than for the SC samples,
at low temperatures there is no reverse: sharper modes deliver a more
intense signal. Even in underdoped cuprates, where there are many reports
of coexisting magnetism and superconductivity, and there is little debate
that the charge carriers and magnetic moments arise from the same electrons,
we are unaware of any reports of such a phenomenon in two-magnon Raman spectra.

We quantify the change in spectra around $T_c$ by integrating the measured
intensities over the two-magnon peak. We adopt three different definitions
of the integration window in order to gain the maximum understanding of the
significance of our results. These are a) integration over the entire range
from 600 to 3920 cm$^{-1}$; b) choosing the window from 1100 to 2600 cm$^{-1}$
to focus on the signal arising from the 110$-$120 meV optical magnons;
c) following the method of Blumberg {\it et al.},\cite{Blumberg} who defined
the integrated two-magnon peak intensity as the excess signal above the minima
on either side of the peak, which defines a type of background-free
contribution between (in our system) 1100 and 2500 cm$^{-1}$. These
integration areas are represented in the respective panels of Fig.~5.

\begin{figure}
\includegraphics[angle=0,width=7.2cm]{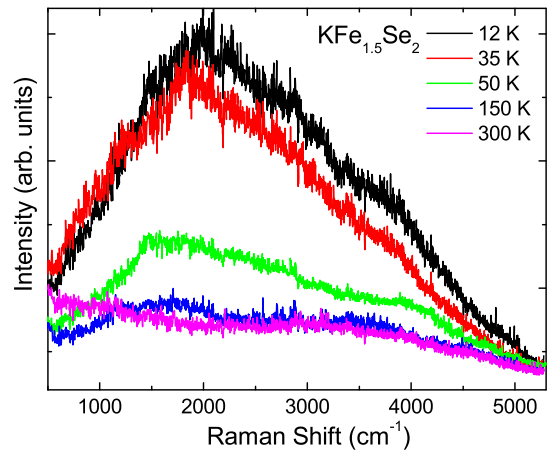}
\caption{(Color online) Temperature-dependence of the Raman intensity
in non-superconducting KFe$_{1.5}$Se$_2$.}
\label{fig4}
\end{figure}

In a further attempt to minimize extraneous effects, we chose to focus on
our sample of Tl$_{0.5}$Rb$_{0.3}$Fe$_{1.6}$Se$_2$ [Fig.~3(b)]. In our sample
of K$_{0.8}$Fe$_{1.6}$Se$_2$ [Fig.~3(a)], the ratio of integrated intensities
between $T_c$ and 9 K does not show a strong effect [by method (b), $I^b(35)
/I^b(9) = 1.02$], but noisy integrated intensity data as a function of
temperature led us to doubt the overall reliability of this sample,
due to its age and to complications with surface light reflections. By
contrast, in our sample of Tl$_{0.5}$K$_{0.3}$Fe$_{1.6}$Se$_2$ [Fig.~3(c)] we
obtain $I^b(35)/I^b(9) = 1.30$, suggesting a massive reduction effect; we
discount this data due to its significantly lower overall intensity and
consequently higher signal-to-noise ratio (Fig.~1). For our sample of
Tl$_{0.5}$Rb$_{0.3}$Fe$_{1.6}$Se$_2$, we measured high-frequency Raman spectra
over a range of narrowly spaced temperature points. The integrated intensities
derived from our data for all three definitions in the previous paragraph are
presented in Fig.~5. All show a steady increase with decreasing temperature
above $T_c$, terminated by a sharp drop occurring essentially at $T_c$. This
drop appears to be continuous, as the rapid opening of the superconducting
gap at $T_c$ acts to remove weight from the two-magnon intensity. The lost
weight then saturates as the gap approaches a constant value at lower
temperatures. Depending on the definition of the integration region, this
loss of weight is a 5$-$10\% effect.

\begin{figure}
\includegraphics[angle=0,width=8cm]{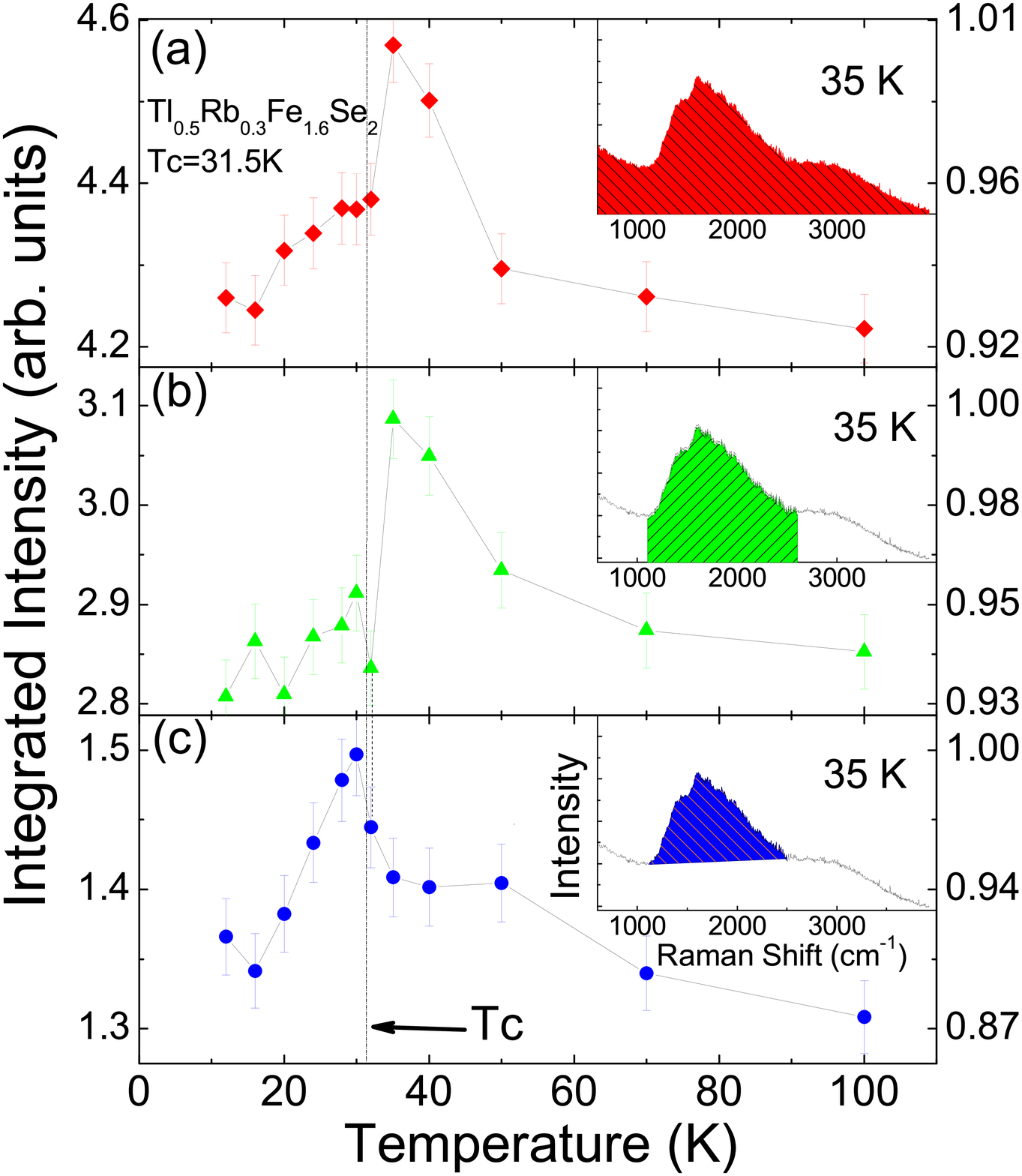}
\caption{(Color online) Temperature-dependence of the integrated scattering
intensity for Tl$_{0.5}$K$_{0.3}$Fe$_{1.6}$Se$_2$. Insets: the shaded regions
represent the integration area used in each panel. (a) Integration over the
full measurement frequency range 600$-$3920 cm$^{-1}$ without background
subtraction. (b) Integration over the primary two-magnon contribution between
1100 and 2600 cm$^{-1}$ without background subtraction. (c) Integration
between the peak minima following Ref.~[\onlinecite{Blumberg}], which
corresponds to a frequency range 1100$-$2500cm$^{-1}$ and the subtraction
of a background term.}
\label{fig5}
\end{figure}

\section{Discussion: competition}

The two-magnon Raman intensity reflects directly a magnetic order. We have
confirmed the presence of superconductivity both before and after our Raman
measurements. The loss of intensity at and below $T_c$ is therefore a clear
statement concerning the coupling and the intrinsic opposition of
superconductivity and antiferromagnetism.

The mutual interaction of magnetism and superconductivity has been difficult
to investigate. While $\mu$SR is the most accurate local probe for estimating
the magnetic volume fraction of a system, the technique gives no information
on the SC state. Neutron diffraction studies have found that the intensity
of the (101) magnetic Bragg peak seems to show a sudden saturation at
$2T_c$.\cite{wbnd} In M\"ossbauer measurements, a change in hyperfine field
below $T_c$ is suggested, but is not detected clearly.\cite{Mossbauer} By
contrast, our results do demonstrate clearly the nature of the relationship.
Because we observe a 5-10\% effect in a system which may be little more than
5-10\% superconducting by volume fraction, we conclude that superconductivity
and magnetism share a strong mutual exclusion.

Within the framework of microscale phase separation discussed in Sec.~I,
one possible scenario for the strong competition we observe between
superconductivity and AF order is a true, microscopic coexistence in the
magnetic phase. This would require that the superconductivity be a bulk
property not only of the paramagnetic (PM) phase observed by NMR and ARPES,
but also of the AF phase. However, band-structure calculations\cite{rlu}
have shown this phase to be a semiconductor, and this scenario would require
that it could become weakly metallic due to non-stoichiometry-induced doping.
The predicted energy gap to the conduction bands makes this possibility
unlikely. Further, it remains to be understood how a system with such
strong magnetism ($T_N = 520$ K) could simultaneously host such strong
superconductivity ($T_c = 32$ K). One year of intensive investigation into
the A$_{x}$Fe$_{2-y}$Se$_2$ materials has not yet revealed any qualitatively
different properties to justify discarding the conventional understanding
concerning generic exclusion of magnetism and superconductivity.

The alternative scenario relies on microscale phase separation. In this
case, the superconductivity is a property of the PM phase alone, but this
phase must be percolating, despite its small volume fraction, for the system
to be a bulk superconductor. The measured volume fractions thus suggest a
system of microscopic magnetic domains, whose disordered boundary regions
are the PM phase. When the latter turn superconducting, they are required
to drive a large fraction of the magnetic volume superconducting by proximity,
reducing the ordered moment by the 5$-$10\% that we observe. Such a physical
proximity effect necessitates large contact areas between the AF and PM
regions, and hence our results confirm that the lengthscale of the phase
separation is required to be nanoscopic. Taken together with all of the
other experimental evidence for nanoscale phase separation (Sec.~I), we
conclude that the microscopic mechanism for the strong competition we
observe between magnetic order and superconductivity is an extensive
grain-boundary proximity effect.

For completeness, we conclude this discussion by excluding two further
scenarios. Because our Raman phonon measurements contain no evidence, in
the form of phonon anomalies, for any type of structural phase transition
near $T_c$,\cite{ZAM1} an explanation for our results in terms of changes
in the magnetism of the AF phase is excluded. Because our experiments probe
many hundreds of atomic layers, and indeed possibly complete domains of the
magnetic structure, they cannot be the consequence of a surface phenomenon.

\section{Conclusion}

In summary, we have measured the high-energy Raman-scattering intensity
in three different A$_{0.8}$Fe$_{1.6}$Se$_2$ superconductors. We find a
robust and reproducible signal in all cases, which can be ascribed to
two-magnon scattering processes involving the optical magnons of the
$\sqrt{5} \times \! \sqrt{5}$ magnetic structure. We have discovered a
striking drop in two-magnon Raman intensity below $T_c$, which has no
precedent in other strongly correlated electronic superconductors. The
magnitude of this competitive effect suggests an almost complete mutual
exclusion of superconductivity and magnetic order due to proximity
effects occurring within a microstructure based on nanoscopic phase
separation.

\acknowledgments

We thank T. Li, Z. Y. Lu, and W. Bao for helpful discussions. This work was
supported by the 973 program under Grant No.~2011CBA00112, by the NSF of
China under Grant Nos.~10874215, 10934008, and 11034012, by the Fundamental
Research Funds for Central Universities, and by the Research Funds of RUC.

\bigskip

\emph{Note added in proof}: after completion of this manuscript, we became
aware of neutron diffraction results from the group of W. Bao.\cite{wbndu}
In exact analogy with our measurements, the magnetic Bragg peak intensities
show a rapid downturn as the temperature is lowered below $T_c$, confirming
the strong competition between the magnetic and superconducting order
parameters.

\end{document}